\documentclass[twocolumn,prb,aps,showpacs,floatfix]{revtex4}  
\usepackage{epsfig,graphics}

\newcommand{\kp}{\mathbf{k_\parallel}}
\newcommand{\Gbar}{\overline{\Gamma}}
\newcommand{\square}{\mathrm{I}}

\begin{document}

\title{Ballistic Spin Injection from Fe into ZnSe and GaAs  \\
  with a (001), (111), and (110) orientation}

\author{O. Wunnicke}
\author{Ph. Mavropoulos}
\author{R. Zeller}
\author{P.H. Dederichs}
\affiliation{Institut f\"ur Festk\"orperforschung, Forschungszentrum
J\"ulich, D-52425~J\"ulich, Germany}

\date{\today}

\begin{abstract}
  We present first-principles calculations of ballistic spin injection
  in Fe/GaAs and Fe/ZnSe junctions with orientation (001), (111), and
  (110). We find that the symmetry mismatch of the Fe minority-spin
  states with the semiconductor conduction states can lead to
  extremely high spin polarization of the current through the (001)
  interface for hot and thermal injection processes. Such a symmetry
  mismatch does not exist for the (111) and (110) interfaces, where
  smaller spin injection efficiencies are found. The presence of
  interface states is found to lower the current spin polarization,
  both with and without a Schottky barrier. Finally, a higher bias can
  also affect the spin injection efficiency.
\end{abstract}


\maketitle


\section{Introduction}


The idea of exploiting both the charge and the spin of an electron in
semiconductor (SC) devices has lead to the growing field of
spintronics. \cite{WA01} Several spintronic devices are already
proposed, \cite{DD90, GS99} but the full potential of spintronics has
still to be discovered. At the moment the essential ingredients for
spintronics, e.g., the injection and detection of a spin polarized
current, the spin transport in SC etc., are only partially achieved in
experiments. In this paper we want to investigate one of these main
challenges: How to create a spin polarized current in a SC? There are
several methods that obtain a high spin polarization, but most are
useful for basic studies rather than applications: illumination of the
SC with circularly polarized light, \cite{HO98, KA98} injection of
electrons from a ferromagnetic scanning tunneling microscope (STM) tip
\cite{AR92, LB01} or from a paramagnetic semiconductor polarized in an
external magnetic field.\cite{FK99, JP00} Better applicable for
devices is the injection from a diluted magnetic
semiconductor,\cite{OY99,Mattana03} but these have the drawback of low
Curie temperatures $T_C$ up to now.  More promising is the injection
from a common ferromagnetic metal, like Fe, into a SC due to the high
$T_C$. In this case it has been shown \cite{SF00} that in the
diffusive limit the spin polarization of the injected current is
vanishingly small, \cite{FH00, GS99, HB99} if the contact between the
FM and the SC is ohmic. This `fundamental obstacle' is traced back to
the conductivity mismatch between both materials. It can be overcome
by either a ferromagnet with nearly 100\% spin polarization at the
Fermi energy, e.g., a half metal, or by inserting a spin polarized
interface resistance at the FM/SC interface, \cite{Ra00, FJ01} e.g.,
an extrinsic tunneling barrier or an intrinsic Schottky barrier. By
means of a Schottky barrier \cite{ZR01, HJ02,HE03} or of an
Al$_2$O$_3$ barrier \cite{MS02} spin injection could be achieved in
experiments.

Here we investigate the spin injection from a ferromagnet (FM) into a
SC in the ballistic regime, {\it i.e.}, the limit where inelastic and
incoherent scattering events are negligible. In this case there is no
scattering in the perfect bulk regions and the whole resistance
results only from the interface region so that there is no
conductivity mismatch. Therefore other effects emerge that are not
present in the diffusive limit: Kirczenow \cite{Ki01} has shown that
some FM/SC interfaces can act as ideal spin filter. This is the case,
if for a direct-gap SC the Fermi surface of the FM projected onto the
two-dimensional Brillouin zone (2DBZ) has a hole at the $\Gbar$ point
for one spin direction. But the material combinations Fe/ZnSe and
Fe/GaAs investigated here, do not fulfill this condition. It has been
pointed out that the symmetry of the FM wave functions is very
important for the spin injection process. \cite{WM02,MW02,ZX03} If
there is a second FM lead to detect the spin polarization of the
current, additional effects like the forming of quantum well states,
Fabry-Perot like resonances and extremely high magnetoresistance ratios
have been reported in a recent theoretical {\it ab-initio}
study. \cite{MW02}

There are two types of calculations for the ballistic spin injection
already published: by means of analytical models using plane waves
\cite{Gr01, HM01, HS01} and by {\it ab-initio} methods. \cite{WM02,
ZX03, MW02, Wu03} While the model calculations obtain only a few
percent of spin polarization, the inclusion of the full band structure
of the FM and the interface region in {\it ab-initio} methods can
result in spin polarizations up to 99\%.
The origin of this high polarization are the different symmetries of
the Fe $d$ states at the Fermi energy for the majority and the
minority spin.

Here we extend our previous work \cite{WM02,MW02} and show more
details of the spin injection process.  In
Sec. \ref{Calculation_details} the details of the calculations and the
investigated Fe/SC systems are given. In Sec. \ref{Sec:Hot} the
ballistic spin injection of hot electrons for the Fe/SC(001), (111),
and (110) interfaces is investigated. It emerges that the (111) and
(110) orientations do not show the large symmetry enforced spin
polarization as the (001) orientation does. In Sec.~\ref{Sec:Thermal}
we report on injection of thermal electrons with and without a
Schottky barrier, addressing also the effect of resonant interface
states. Sec.~\ref{Sec:Bias} gives an approximation to the effect of a
higher bias voltage (out of the linear-response regime). The paper
closes with a summary of the results in Sec. \ref{summary}.

\section{Calculation details}\label{Calculation_details}

The investigated heterostructures consist of a half-space of ideal
bulk Fe and one of the ideal bulk SC (either ZnSe or GaAs). Their
properties are described by the surface Green's function determined by
the decimation technique. \cite{TD97} Between the half-spaces there
are several monolayers (MLs) of the interface region, where the atomic
potentials are allowed to deviate from their bulk values. This
interface region consists of four MLs of Fe and two MLs of SC for the
(001) orientation and appropriate numbers of MLs for the (111) and
(110) orientations. The corresponding potentials are obtained
self-consistently from a Fe/SC/Fe junction geometry by the screened
Korringa-Kohn-Rostoker (KKR) Green function
method.\cite{WZ97,Papanikolaou02} It is assumed that the SC lattice is
matched to the bulk Fe, resulting in an fcc SC lattice constant double
the lattice constant of bcc Fe $d_\mathrm{SC} = 2 d_\mathrm{Fe} =
5.742$\AA. The experimental lattice constant of Fe is used. This means
that the SC is stretched compared to the bulk lattice constant by
1.3\% and 1.6\% for ZnSe and GaAs, respectively, resulting in a
slightly smaller energy gap. In order to describe the SC accurately,
two empty spheres per unit cell are introduced to account for the open
space in the geometry of the zinc-blende structure. A perfect
two-dimensional translation symmetry in the whole system is assumed
and so the in-plane component $\kp$ of the $\mathbf{k}$ vector is
conserved while crossing the interface. The $z$ axis is always assumed
to be the growth direction, {\it i.e.}, standing perpendicular on the
interface. Also the conductance $G$ is calculated by a two-dimensional
Fourier transformation in dependence of $\kp$.

The calculations of the ground state properties are performed within
the density-functional theory in the local density approximation (LDA)
for the exchange and correlation terms. It is well known that it
underestimates the energy gap in SC by around one half. Otherwise it
gives very accurate FM and SC bands and is well suited for the
considered problem. The potentials are described within the
atomic-sphere approximation, and the wave functions are expanded in
angular momentum up to a cutoff of $\ell_{max} = 2$ for
self-consistency and $\ell_{max}=3$ for the conductance calculations.
Spin-orbit coupling and spin-flip scattering are not included in the
calculations. This could have an effect in case of injection into the
SC valence band, where the spin-orbit interaction is stronger, but not
in the case of the valence band which has mainly $s$ character and
negligible spin-orbit coupling.

The conductance $G$ is calculated by the Landauer formula.
\cite{La57} $G$ is expressed as a sum of the transmission
probabilities through the scattering region (here the Fe/SC interface)
over all available conducting channels. These channels are enumerated
by the spin orientation $\sigma = \uparrow$ or $\downarrow$, the bands
$\nu, \nu'$ at the Fermi energy of the left and right lead,
respectively, and the $\kp$ vector. The spin-dependent conductance,
normalized to the area of the two-dimensional unit cell, reads
\begin{equation}
G^{\sigma} = \frac{e^2}{h} \frac{1}{A_\mathrm{2DBZ}} \sum_{\nu, \nu'} 
\int \limits_\mathrm{2DBZ} d^2 k_\parallel  \
T^\sigma_{\nu, \nu'} (\kp) \,
\label{eq.landauer}
\end{equation}
with $A_\mathrm{2DBZ}$ the area of the two-dimensional Brillouin zone.
Since the $\kp$ vector and the spin $\sigma$ of the electron are
conserved in our calculation, they are the same for the incoming and
the outgoing states. Here the Landauer formula is evaluated by a
Green's function formalism introduced by Baranger and
Stone. \cite{BS89} Details are presented elsewhere. \cite{MPxx} The
spin polarization $P$ of the current is defined by the spin dependent
conductances in both spin bands
\[
P = \frac{G^\uparrow - G^\downarrow} {G^\uparrow + G^\downarrow},
\]
where $G^\uparrow$ and $G^\downarrow$ is the conductance of the
majority and the minority electrons, respectively.


\section{Hot electron injection process \label{Sec:Hot}}
To investigate the effects of the symmetry of the Fe bands, we
calculate first the hot injection of electrons, {\it i.e.}, of Fe
electrons well above the Fermi level $E_F$, which falls in the gap of
the SC. For most of the calculations we restrict the discussion to the
$\Gbar$ ($\kp=0$) point for simplicity, {\it i.e.}, we consider only
electrons with perpendicular incidence on the interface. This is
motivated by the fact that in most spin injection experiments the
Fermi energy in the SC is only some tens of meV above the conduction
band minimum resulting in a very small Fermi sphere around the
$\Gamma$ point.

\subsection{The Fe/SC(001) orientation}\label{001}

First we investigate the spin injection in Fe/SC junctions
grown in the (001) orientation. The (001) direction has the highest
symmetry and is therefore the best candidate for a {\it symmetry enforced
high spin polarization of the injected current,\cite{WM02} {\it i.e.},
almost total reflection of the minority spin electrons due to the
symmetry mismatch between the SC and minority spin Fe bands.}

For the investigation of the bands that are available in Fe and in the
SC at the $\Gbar$ point, it is important to note that the in-plane
common two-dimensional unit cell is double in area than the one of
bulk bcc Fe (containing two inequivalent Fe atoms), and the common
2DBZ has half the area of the one of bcc Fe. Thus additional bands are
backfolded into the 2DBZ and the number of bands at $\Gbar$
increases. A description of the backfolding is given in the Appendix.

\begin{table}
\caption{Symmetry properties of energy bands in Fe [001] ($C_{4v}$
  symmetry group) and their local orbital analysis. The backfolded
  bands are indicated by (backf.). The last column shows the coupling
  to the SC $\Delta_1$ conduction band ($C_{2v}$ symmetry group) based
  on symmetry compatibility relations and local orbital form ($d$
  orbitals are localized and expected to couple poorly; $p_z$ orbitals
  are extended into the SC and expected to couple well).}
\label{Table:Sym}
\begin{tabular}{lll}
\hline
Rep. in Fe & Orbital analysis  & Coupling to SC $\Delta_1$ \\
\hline
$\Delta_1$ & $s; p_z; d_{z^2}$ & good                      \\
$\Delta_{2'}$ & $d_{xy}$       & poor                      \\
$\Delta_2$ & $d_{x^2-y^2}$     & none (orthogonal)         \\
$\Delta_5$ & $p_x; p_y; d_{xz}; d_{yz}$ & none  (orthogonal)           \\
D$_1$ (backf.) & $s; p_x; d_{yz}; d_{x^2-y^2}-d_{z^2}$ & poor       \\
D$_2$ (backf.) & $d_{x^2-y^2}+d_{z^2}$  & poor                      \\
D$_3$ (backf.) & $p_y+p_z; d_{xy}+d_{zx}$ & poor              \\
\hline
\end{tabular}
\end{table}

\begin{figure} [tb]
  \begin{center}
    \includegraphics[width=3.3in,angle=0]{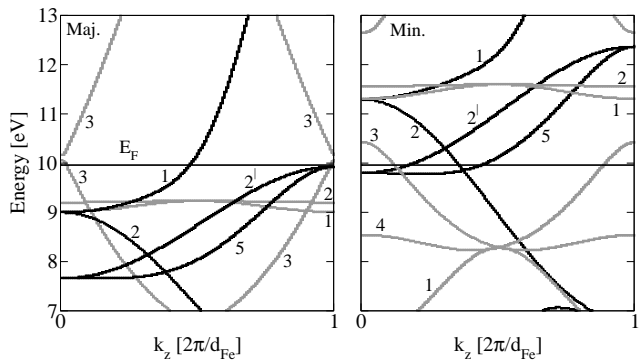}
    \caption{Band structure of Fe(001) at the $\Gbar$ point of the
    2DBZ (black lines represent the bulk $\Gamma$-H bands and gray
    lines the backfolded N-P-N bands). The left panel shows the
    majority and the right one the minority spin band structure. The
    Fermi energy is shown by a horizontal line. The numbers denote the
    symmetry of the bands along the $\Delta$ direction ($\Gamma$-H)
    and the D direction (N-P-N). \cite{CW77}}
    \label{band_structure_Fe_001} \end{center}
\end{figure}

\begin{figure} [tb]
  \begin{center}
    \includegraphics[width=3.3in,angle=0]{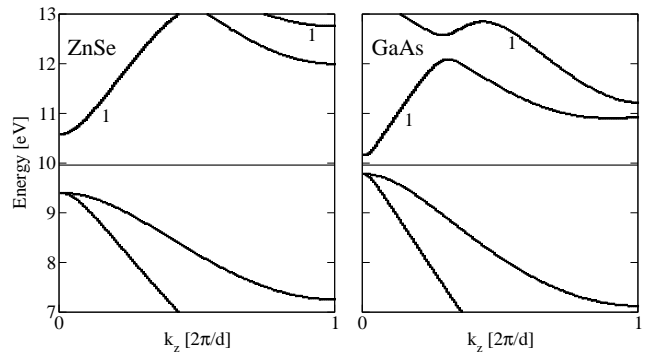}
    \caption{Band structure at the $\Gbar$ point of the 2DBZ for ZnSe (001)
      (left panel) and GaAs (001) (right panel). The numbers denote the
      symmetry of the bands along the $\Delta$ direction.
      \cite{WK81} $d$ is the SC lattice constant.}
    \label{band_structure_SC_001}
  \end{center}
\end{figure}

The Fe band structure at the $\Gbar$ point along $k_z$ ($\Gamma$-H,
{\it i.e.} the $\Delta$ direction), including the backfolded bands, is
shown in Fig.~\ref{band_structure_Fe_001} for both spins. This
direction has in bulk Fe a $C_{4v}$ symmetry, just as the Fe (001)
surface.
In the 2DBZ of the SC no bands are backfolded to the $\Gbar$
point. The ZnSe and GaAs band structures along $\Gamma$-X (the
$\Delta$ direction) are shown in Fig.~\ref{band_structure_SC_001}. In
the SC this direction has a $C_{2v}$ symmetry (a subgroup of $C_{4v}$),
as the SC (001) surface.
Therefore the Fe/SC(001) interface has also the reduced $C_{2v}$
symmetry. As a result, some bands $d$ that are mutually orthogonal in
bulk Fe are able to couple to each other at the interface. 

Now it is important to examine which bands of Fe are allowed to couple
to the $\Delta_1$ conduction band of the SC which transforms fully
symmetrically under the rotations of $C_{2v}$; this can be found by
the compatibility relations between the groups $C_{4v}$ and $C_{2v}$.
Note that the symmetry notation of Fe and the SC refers to the
different groups, $C_{4v}$ for Fe and $C_{2v}$ for the SC. At $E_F$
and above the available Fe bands along the $\Delta$ direction (black
lines in Fig.~\ref{band_structure_Fe_001}) and their symmetry and
coupling properties are given in Table~\ref{Table:Sym}. Practically,
the only ones that are able to couple well are the $\Delta_1$ bands
containing the extended $s$ and $p_z$ orbitals, as well as the
$d_{z^2}$ orbitals pointing into the SC, while the $\Delta_{2'}$ bands
couple only poorly because they consist of more localized $d_{xy}$
orbitals pointing in-plane.  At $E_F$, and up to 1.2~eV above, the
former bands exist only for majority spin, allowing transmission,
while for minority spin the latter bands co-exist with the $\Delta_2$
and $\Delta_5$ bands.

\begin{figure} [tb]
  \begin{center}
    \includegraphics[width=3.1in,angle=0]{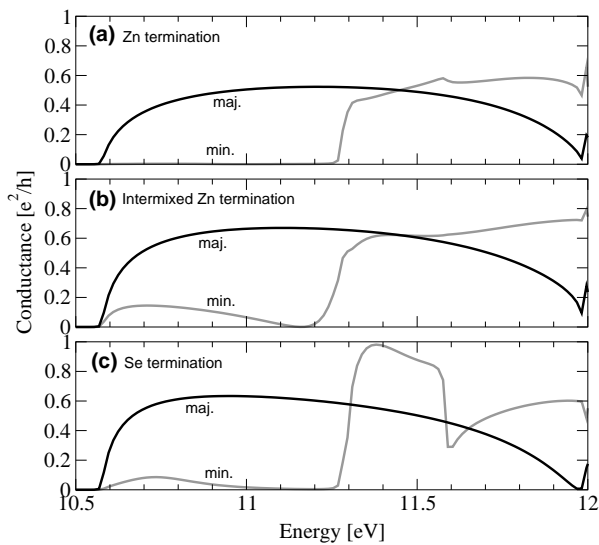}
    \caption{Hot injection of electrons in Fe/ZnSe(001) with a Zn
      terminated (a), an intermixed Zn terminated (b) and a Se
      terminated (c) interface. The black line shows the majority and
      the gray line the minority spin. The conductance is evaluated at
      the $\Gbar$ point for simplicity.}
    \label{hot_injection_Fe_ZnSe_001}
  \end{center}
\end{figure}
\begin{figure} [tb]
  \begin{center}
    \includegraphics[width=3.1in,angle=0]{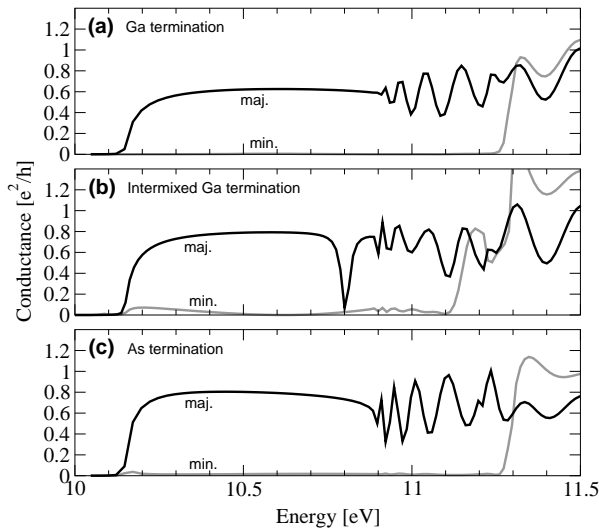}
    \caption{Caption as in Fig.~\ref{hot_injection_Fe_ZnSe_001} but
      for Fe/GaAs(001) with a Ga terminated (a), an intermixed Ga
      terminated (b) and an As terminated interface. Black lines refer
      to majority and gray lines to minority spin directions.}
    \label{hot_injection_Fe_GaAs_001}
  \end{center}
\end{figure}

In Fig.~\ref{hot_injection_Fe_ZnSe_001} the calculated conductance for
the hot injection process for Fe/ZnSe(001)
and in Fig.~\ref{hot_injection_Fe_GaAs_001} for Fe/GaAs(001) are
shown. In (a) the Zn and Ga terminated and in (c) the Se and
As terminated interfaces are shown. The intermixed interfaces (b) will
be discussed later. 
As it was already discussed above, the coupling of the minority
$\Delta_{2'}$ band is much weaker than the coupling of the majority
$\Delta_1$ band to the $\Delta_1$ conduction band of the SC. Thus the
$\Delta_{2'}$ states are nearly totally reflected at the interface.
This results for all three terminations in a very high symmetry
enforced spin polarization. Around 1.3~eV above the Fermi energy (at
11.3~eV) also a $\Delta_1$ band is available in the minority channel.
At this energy the conductance in the minority channel rises
drastically showing that the high spin polarization at lower energies
originates precisely from the absence of the $\Delta_1$ Fe band in the
minority channel. Also the flat backfolded bands of D$_1$ and D$_2$
symmetries give a small contribution to the conductance in the small
energy window of around 0.3eV width, centered around $11.4\ $eV. At
these energies in the ZnSe half-space there is only one conduction
band available at the $\Gbar$ point, limiting the maximum conductance
to 1~$e^2/h$. In the case of the GaAs half-space there are three
conduction bands and so the conductance could rise up to a maximal
value of 3 $e^2/h$.

Finally the conductance can be seen to decrease for small conduction
band energies and even vanishes at the conduction band minimum of the
SC. This can be explained qualitatively by the vanishing group
velocity, so that no current can be transported away from the
interface.


\subsection{The Fe/SC(111) interface}\label{111}

In this section the spin injection for the Fe/SC junctions grown in
the (111) orientation is investigated. In the Fe half-space the [111]
direction has a hexagonal symmetry with one atom per unit cell. But in
the SC half-space this direction has only a three-fold rotational
axis. In each ML there is either a cation or an anion (or one of the
two types of empty spheres). Also there are a two kinds of possible
geometric terminations: one where the terminated SC atom is singly and
one where it is triply coordinated to the Fe atoms.

\begin{figure} [tb]
  \begin{center}
    \includegraphics[width=3.0in,angle=0]{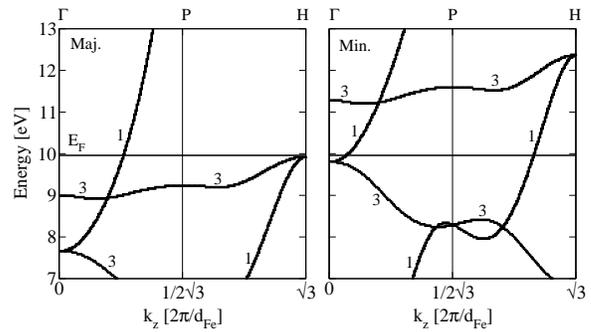}
    \caption{Band structure of Fe(111) at the $\Gbar$ point of the 2DBZ
      (corresponding to the bulk band along the $\Gamma$-P-H symmetry
      line). The left panel shows the majority and the right panel the
      minority spin. The Fermi energy is shown by a horizontal line.
      The numbers denote the symmetry of the corresponding $\Lambda$-
      (between $\Gamma$ and P) and F- (between P and H) symmetry
      direction.  \protect{\cite{CW77}}}
    \label{band_structure_Fe_111}
  \end{center}
\end{figure}

In Fig.~\ref{band_structure_Fe_111} the band structure of Fe(111) at
the $\Gbar$ point in the [111] direction is shown.
In the SC(111) 2DBZ no bands are backfolded to the $\Gbar$ point and
the only available band corresponds to the bulk band along the
$\Gamma$-L high symmetry line.

\begin{figure} [tb]
  \begin{center}
    \includegraphics[width=3.0in,angle=0]{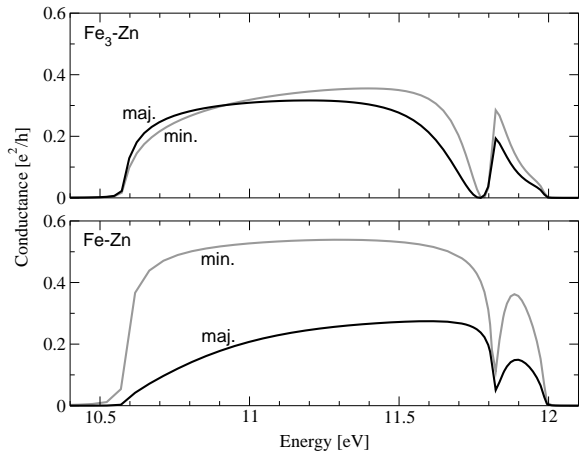}
    \caption{Hot injection of electrons in Fe/ZnSe(111) for a Zn terminated
      interface, where the Zn atoms are either triply coordinated
      (upper plot) or singly coordinated (lower plot) with respect to
      the interface Fe atoms. The black lines refer to the majority
      and the gray lines the minority spin direction. The conductance
      is evaluated at the $\Gbar$ point for simplicity.}
    \label{hot_injection_Fe_ZnSe_111}
  \end{center}
\end{figure}

The band structure plot for Fe(111) shows important differences
compared to the (001) orientation: At the Fermi energy in both spin
channels there are bands with the same $\Lambda_1$ symmetry and there
cannot be any symmetry enforced spin polarization. All eventually
obtained differences in the conductances for both spin directions are
due to the different coupling of the Fe bands to the SC.  This
coupling is believed to be sensitive to the exact interface properties
like the termination, lattice relaxations etc.  The calculated results
for the hot injection process in Fe/ZnSe(111) with a Zn terminated
interface are shown in Fig.~\ref{hot_injection_Fe_ZnSe_111}.
Also for this orientation the conductance is shown at the $\Gbar$
point as before. 

In both the singly and the triply-coordinated Zn terminations the
obtained spin polarization is much smaller than in the (001)
interface, demonstrating the lack of a symmetry enforced spin
filtering. Above 12.0eV there is no ZnSe band exactly at the $\Gbar$
point in the [111] direction, and so no propagating states are
available.


\subsection{The Fe/SC(110) interface \label{110}}

This section deals with Fe/SC(110) oriented junction.  This is the one
with the lowest symmetry considered in this work. From this point of
view a high symmetry-enforced spin polarization is most improbable.
Also the surface unit cell in the SC half-space contains all four
kinds of atoms (cation, anion and both types of vacancies) and is
larger than the (001) and the (111) unit cells. By matching the two
dimensional translational symmetry, also the Fe surface unit cell has
to contain four Fe atoms. This results in a very small 2DBZ for this
orientation leading even in the SC half-space to backfolded bands at
the $\Gbar$ point.

\begin{figure} [tb]
  \begin{center}
    \includegraphics[width=3.3in,angle=0]{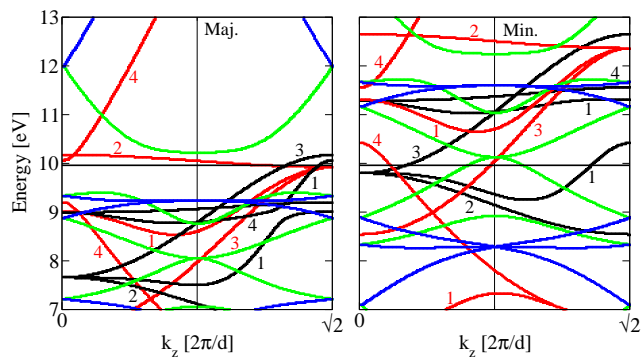}
    \caption{(color online) Band structure of Fe(110) at the $\Gbar$ point of the
      2DBZ. The different colors denote the different backfolded bands
      corresponding to the following bulk bands: black: $\Gamma$-N,
      red: $N$-$H$, green: $\mathbf{b}_1$ applied, blue:
      $\mathbf{b}_1$ and $\mathbf{b}_2$ applied (see the Appendix).
      The left plot shows the majority and the right plot the minority
      spin states. The Fermi energy is shown by an horizontal line.
      The numbers give the symmetry of the corresponding bulk
      band,\cite{CW77} if possible.}
    \label{band_structure_Fe_110}
  \end{center}
\end{figure}

The band structure of Fe (001) at the $\Gbar$ point with the different
backfolded bands are shown in Fig.~\ref{band_structure_Fe_110}.
In the SC(110) half-space there is also a backfolded band at the
$\Gbar$ point (created by applying $\mathbf{b}_1$ in
Fig.~\ref{bz_110}). Because of this the maximum conductance per spin
direction is $2 e^2/h$. 

\begin{figure} [tb]
  \begin{center}
    \includegraphics[width=3.0in,angle=0]{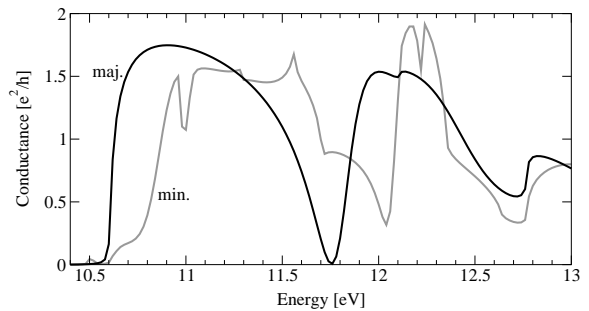}
    \caption{Hot injection of electrons in Fe/ZnSe(110). The black
      line show the majority and the gray line the minority spin
      conductance. Since all four
      different atoms (Zn, Se, and two vacancies) are located in each
      ML, there is only one possible termination. The
      conductance is evaluated at the $\Gbar$ point for simplicity.}
    \label{hot_injection_Fe_ZnSe_110}
  \end{center}
\end{figure}

The variety of backfolded bands and the reduced symmetry of the [110]
direction do not allow for a simple symmetry-based discussion.
Basically, for the (110) oriented heterojunctions there is no symmetry
enforced spin polarization. This is also shown in
Fig.~\ref{hot_injection_Fe_ZnSe_110}, where the conductances at the
$\Gbar$ point for the hot injection process are presented.

From the above we conclude that the best candidate for a
symmetry-enforced high current spin polarization is the (001)
interface. Thus, in what follows, we restrict our study to this case.


\subsection{Hot injection process $\kp$ resolved}

\begin{figure} [tb]
  \begin{center}
    \includegraphics[width=4.0in,angle=270]{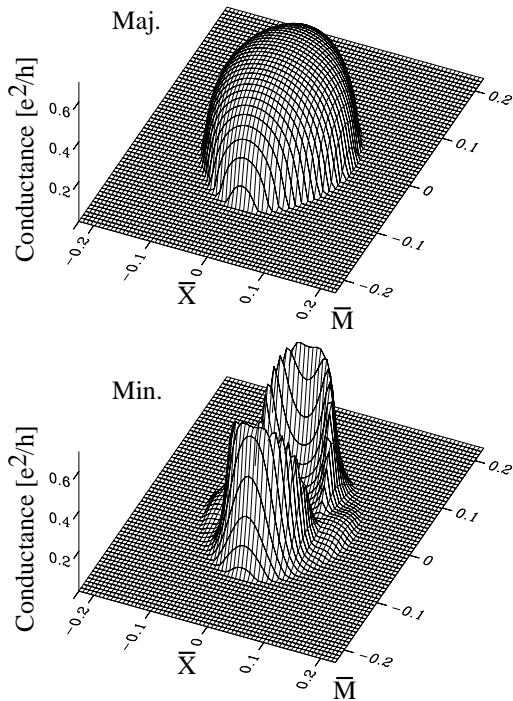}
    \caption{Conductance in the hot injection process in Fe/ZnSe
      (001) (Zn termination), resolved in $\kp$. The upper plot shows
      the conductance of the majority and the lower one for the
      minority spins. Only one tenth of the 2DBZ around the $\Gbar$
      point is shown. The Fermi energy lies 500~meV above the
      conduction band minimum in the semiconductor half-space. The
      reciprocal vectors are in units of $2\pi/d_{\mathrm{SC}}$.}
    \label{hot_injection_k_parallel} \end{center}
\end{figure}

In the hot injection process the restriction to the $\Gbar$ point was
motivated by the fact that in most applications the Fermi level is
positioned only slightly in the conduction band resulting in a very
small Fermi sphere around the $\Gbar$ point. More correctly the
conductance should be integrated over the whole 2DBZ, because also
states away from the $\Gbar$ point are populated for higher injection
energies. Therefore in Fig.~\ref{hot_injection_k_parallel} the $\kp$
resolved conductances are shown for the majority and minority
electrons for the Zn-terminated Fe/ZnSe (001) interface.
The energy of the injected electrons is 500~meV above the conduction
band minimum. In this case the polarization at the $\Gbar$ point is
99.7\%. If the conductance is integrated over the 2DBZ the
polarization is reduced to 39.5\%. The reason is a higher conductance
in the minority band for states away from the $\Gbar$ point. The
symmetry arguments discussed above are valid only at this high
symmetry point. For states $\kp \ne 0$ also other Fe minority states
are allowed to couple to the SC $\Delta_1$ conduction band and so the
conductance rises. We conclude that the high symmetry-enforced spin
polarization can be realized only for energies up to a few tens of meV
above the conduction band edge.

In the conductance plot (Fig.~\ref{hot_injection_k_parallel}) for the
minority electrons the reduced $C_{2v}$ symmetry of the Fe/SC(001)
interface can be directly seen. The majority conductance behaves like
free electrons traveling across a potential step and is practically
unaffected by the reduced symmetry. The circularly symmetric form of
$G^{\uparrow}(\kp)$ and the twofold-symmetric form of
$G^{\downarrow}(\kp)$ can be understood in terms of the form of the
Bloch functions for small deviations from $\kp=0$ in the same manner
as described in Ref.~\onlinecite{MW02} for the Fe/SC/Fe (001)
junctions.


\subsection{The intermixed interface}

By means of an {\it ab-initio} study \cite{EL02} it has been found
that the interface of Fe/GaAs(001) with a Ga terminated interface is
energetically more stable, if Fe atoms diffuses into the vacancy sites
of the first Ga ML. In case of an As terminated interface the abrupt
interface is found to be more stable than the intermixed one. The
former case is accompanied by rather large lattice distortions which
are not taken into account in our calculations. Although we do not
know of analogous studies for Fe/ZnSe(001) interfaces we have also
performed the intermixed interface in case of a Zn termination. The
results are shown in Fig.~\ref{hot_injection_Fe_ZnSe_001}(b) and
\ref{hot_injection_Fe_GaAs_001}(b) for the hot injection process with
an intermixed Zn and Ga interface, respectively. The still high spin
injection polarization can be explained by the fact that the
intermixed Fe atoms do not change or reduce the $C_{2v}$ symmetry of
the interface and so the symmetry arguments are still valid. This is a
different effect than interface roughness that reduces the spin
polarization drastically. \cite{ZX03,SH02} Solely the coupling of the
Fe states to the SC are changed. In both cases the coupling in the
minority band is enhanced in comparison to the atomically abrupt
interfaces shown in Fig.~\ref{hot_injection_Fe_ZnSe_001}(a) and
\ref{hot_injection_Fe_GaAs_001}(a).

\section{Thermal injection process \label{Sec:Thermal}}

In this section the injection of thermal electrons is investigated.
This is achieved by lowering the potentials in the SC half-space by a
rigid shift, so that the Fermi energy falls slightly above the
conduction band minimum (here around 10 meV). All other potentials are
kept in their ground state position, in particular also the potentials
of the two SC monolayers in the interface region. This shifting
simulates the effect of n-doping or an applied gate voltage in a
field-effect-transistor device. The electrons are injected at the
$E_F$ of the Fe half-space directly into the conduction band. Since
the Fermi energy is only some tens of meV above the conduction band
minimum and since effective masses of ZnSe and GaAs are small, the
resulting Fermi wave vector in the SC half-space is very small (around
one hundredth of the distance to the boundary of the 2DBZ). So we
present here the conductance only for the $\Gbar$ point. The evaluated
polarization changes only slightly, if the conductances are integrated
over the whole 2DBZ.

\subsection{Thermal injection without Schottky barrier}

\begin{figure} [tb]
  \begin{center}
    \includegraphics[width=3.0in,angle=0]{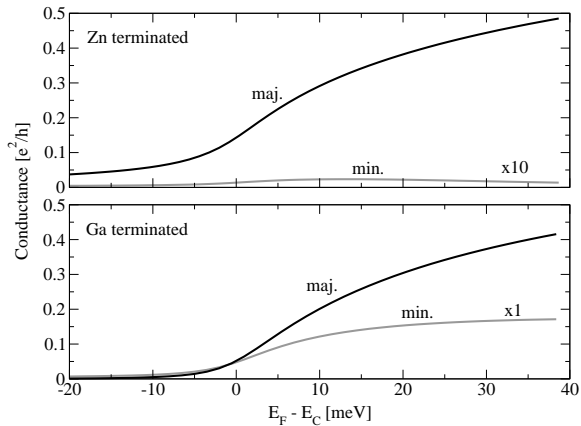}
    \caption{Thermal injection of electrons in Fe/ZnSe (upper plot)
      and Fe/GaAs (001) (lower plot) with an intermixed Zn or Ga
      interface, respectively. The black line shows the majority and
      the gray line the minority spin. The conductance is evaluated at
      the $\Gbar$ point. Note that the minority conductance is
      enlarged by a factor of 10 in case of the Zn termination.}
    \label{thermal_injection_Fe_SC_001_diffused}
  \end{center}
\end{figure}

\begin{figure} [tb]
  \begin{center}
    \includegraphics[width=3.0in,angle=0]{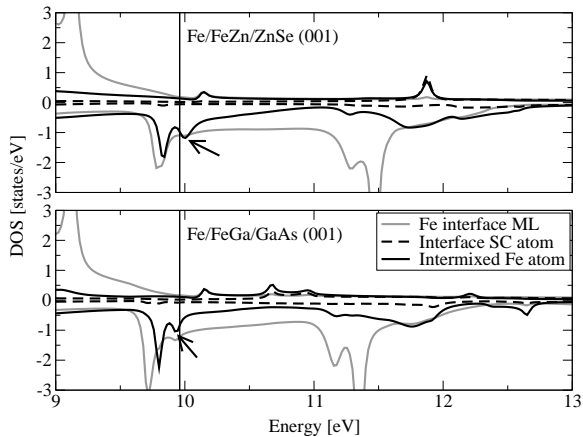}
    \caption{Density-of-states (DOS) for the thermal injection process
      in Fe/ZnSe (001) (upper plot) and Fe/GaAs (001) (lower plot) 
      with an intermixed Zn or Ga interface. Arrows indicate the
      resonance at the intermixed Fe atom.}
    \label{DOS_Fe_SC_001_diffused}
  \end{center}
\end{figure}

The results for Fe/ZnSe and Fe/GaAs(001) are discussed in
Ref.~\onlinecite{WM02} and show a very high spin polarization of more
than 97\% for ZnSe and practically 100\% for GaAs (001). Similar
results are found for tunneling in the Fe/SC/Fe (001)
junctions.\cite{MW02} Here we add the thermal injection with an
intermixed Zn- and Ga-terminated interface
(Fig.~\ref{thermal_injection_Fe_SC_001_diffused}).
In the case of an intermixed Zn interface nearly the same conductances
are obtained as for the abrupt one (Fig.~3 in
Ref.~\onlinecite{WM02}). But a drastic increase of the minority
conductance can be seen for the intermixed Ga interface, resulting in
a much smaller spin polarization. The reason for this increase is an
interface resonance localized at the intermixed Fe atoms. This
can be seen in the density-of-states (DOS) for this heterojunction
shown in Fig.~\ref{DOS_Fe_SC_001_diffused} for the intermixed Zn and
Ga interface. The arrows indicate the position of the Fe resonance. In
the case of the intermixed Zn interface this resonance lies slightly
above the $E_F$ and give rise to the small maximum in the minority
conductance at $E_F - E_C = 15\ $meV seen in
Fig.~\ref{thermal_injection_Fe_SC_001_diffused}.
In the DOS for the intermixed Ga interface this resonance lies nearly
on the Fermi energy and is much narrower, so that it has a strong
influence on the minority conductance. Since the energy position of
this resonance is believed to depend strongly on the interface
properties, e.g., lattice relaxations, further investigations are
needed for the exact energy position. If the injection process is
evaluated at only 0.3eV higher energies, the influence of this
resonance is negligible and the normal high spin polarization is
restored. However, for a higher bias one must integrate over an energy
interval (see below) and thus include resonant interface states that
lie close to, but not at, $E_F$.

\subsection{Thermal injection through a Schottky barrier}

In theoretical studies for strongly diffusive transport\cite{Ra00,
FJ01} it is shown that the negligible spin polarization due to the
conductivity mismatch between the ferromagnetic metal and the
semiconductor\cite{SF00} can be overcome by a spin dependent tunneling
barrier at the interface. Such a barrier could be the Scottky barrier
created at the interface. Although the approaches in
Refs.~\onlinecite{Ra00,FJ01,SF00} assume diffusive transport,
tunneling barriers can be described in the ballistic regime.

In this section the thermal injection process is extended to the case
where a Schottky barrier at the interface is included. The potentials
for this junctions are the same as in the thermal injection process,
but with an inserted tunneling barrier at the interface.  The barrier
is modeled by a rigid shift of each SC atom potential, constant
within each monolayer. It starts at the third SC layer from the
interface, and increases linearly in magnitude with distance from the
interface. In this way the SC bands are gradually shifted downward.
Finally, at the end of the barrier $E_F$ lies 10~meV above the
conduction band edge (and the shift is not changed from then on),
while at the interface $E_F$ lies in the middle of the gap. The
barrier thickness is then varied between 0 and 140 MLs.

\begin{figure} [tb]
  \begin{center}
    \includegraphics[width=8.5cm,angle=0]{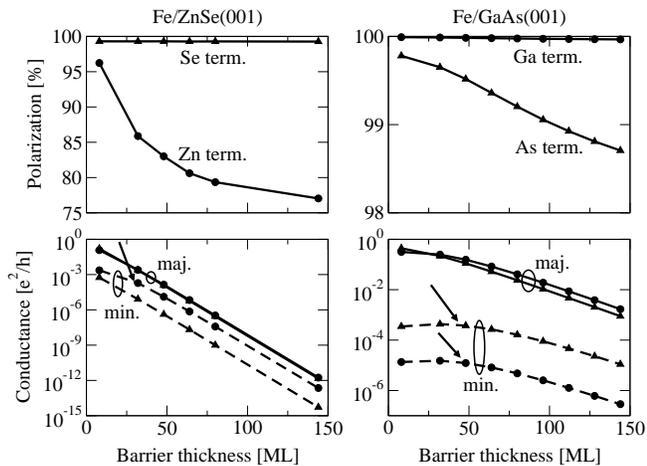}
    \caption{Influence of a smooth Schottky barrier on the spin
    dependent conductance (lower plots) and the spin polarization
    (upper plots) for Fe/ZnSe(001) (left panel) and Fe/GaAs(001)
    (right panel). Circles show the conductance for the Zn or Ga and
    triangles for the Se or As termination. The conductance is
    evaluated at the $\Gbar$ point. Arrows indicate the influence of
    the resonant interface state causing a non-exponential decay of
    the conductance for moderate barrier thickness. In Fe/ZnSe(001)
    the lines for the majority conductances lie on top of each other.}
    \label{tunneling_barrier_Fe_SC_001} \end{center}
\end{figure}

As shown in Ref.~\onlinecite{WM02} a tunneling barrier at the Fe/SC
(001) interface gives also a high spin polarization. Only in case of a
resonant interface state near $E_F$, especially in Fe/ZnSe(001) with a
Zn terminated interface, the polarization is reduced. Here we will
discuss the effect in more detail. In
Fig.~\ref{tunneling_barrier_Fe_SC_001} the conductance and the
polarization in dependence of the barrier thickness is shown for
Fe/ZnSe and Fe/GaAs(001), respectively.
Due to the small Fermi wave vector, the conductance is again analyzed
only at the $\Gbar$ point. The polarizations agree with the ones
correctly integrated over the 2DBZ within 5\%. 

\begin{figure} [tb]
  \begin{center}
    \includegraphics[width=3.0in,angle=0]{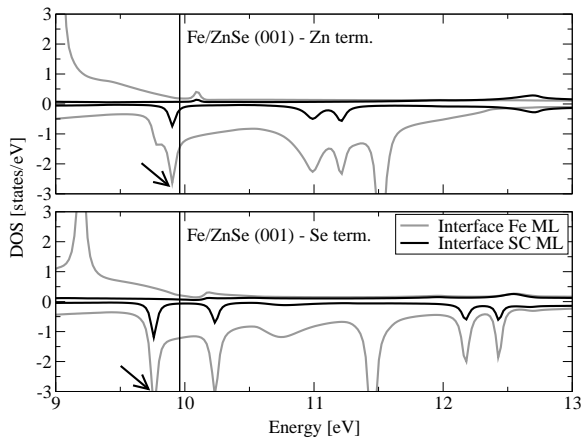}
    \caption{DOS at the $\Gbar$ point for Fe/ZnSe(001) with a Zn
      (upper plot) and a Se terminated interface (lower plot). Gray
      lines show the DOS of the Fe interface ML and black lines the
      DOS of the Zn or Se interface ML.  The vertical line indicate
      the position of the Fermi energy. All potentials are in the
      ground state position. Arrows indicate the peak in the DOS due
      to the resonant interface state.}
    \label{dos_Fe_ZnSe_001}
  \end{center}
\end{figure}

Except for the Se terminated Fe/ZnSe(001) interface, all terminations
show the influence of a resonant interface state in the minority band,
resulting in an higher conductance for the minority spins with thicker
barriers. This leads to a reduction of the spin polarization. For
Fe/GaAs(001) this effect is also visible in the conductance but is
negligible for the spin polarization due to the much larger difference
of the conductance for both spins than in Fe/ZnSe. To discuss this
effect in more detail, in Fig.~\ref{dos_Fe_ZnSe_001} the DOS at the
$\Gbar$ point is shown for Fe/ZnSe(001). The DOS for Fe/GaAs(001) is
qualitatively comparable for this purpose.
In the DOS an interface state is visible near the Fermi energy in the
minority spin channel. This state has $\Delta_1$ symmetry and lies in
the energy region of the $\Delta_1$ hybridization gap of the Fe
minority band structure. It can penetrate well a Schottky barrier of
moderate thickness since the evanescent wave with the longest decay
length is of the same $\Delta_1$ symmetry.\cite{MP00} In the Fe
half-space this interface state is normally also evanescent since
there are no propagating states it can couple to. But due to the
reduced symmetry of the interface the $\Delta_1$ interface state
couples weakly to the Fe $\Delta_{2'}$ band and becomes resonant. In
the case of the Se terminated Fe/ZnSe(001) interface the interface
state lies further away from $E_F$ and practically does not contribute
to the conductance.

The potentials used for the DOS in Fig.~\ref{dos_Fe_ZnSe_001} are the
ground state potentials, {\it i.e.}, $E_F$ is in the middle of the gap
in the SC half-space and no Schottky barrier is inserted. If now a
Schottky barrier is introduced into the junction, the system and the
resonant interface state are slightly distorted, because the
$\Delta_1$ interface state interacts with the $\Delta_1$ conduction
band of the semiconductor. This results in a slight downshift in
energy of the resonant interface state away from $E_F$ for smaller
Schottky barriers. The reduction of the spin polarization with thicker
Schottky barriers can thus be explained by a shift of the resonant
interface state towards $E_F$.

\begin{figure} [tb]
  \begin{center}
    \includegraphics[width=8.5cm,angle=0]{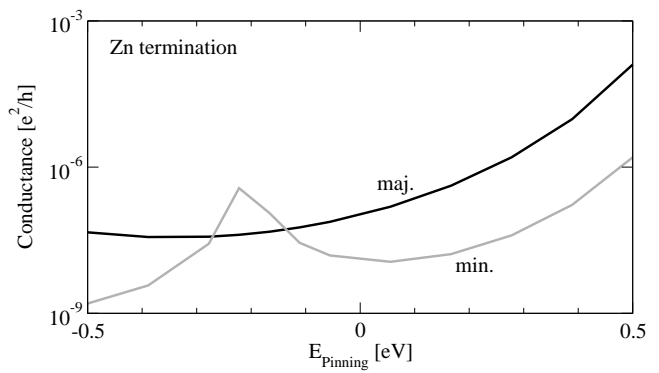}
    \caption{Influence of the pinning position of the Fermi energy
      relative to the middle of the gap on the conductance in
      Fe/ZnSe(001) junctions with a Zn terminated interface. The
      Schottky barrier is 80 ML thick. The black line shows the
      conductance for the majority and gray line of the minority
      spin. Negative energy means a pinning position of the Fermi
      energy near the valence band and positive ones near the
      conduction band. The conductance is evaluated at the $\Gbar$
      point. The minority-conductance peak at $-0.2\ $ eV reflects the
      influence of the interface state.}
    \label{interface_states_Fe_ZnSe_001} \end{center}
\end{figure}

This effect can also be seen by changing the pinning position of the
Fermi energy relative in the SC gap at the interface (the potentials
in the interface region are kept fixed). In
Fig.~\ref{interface_states_Fe_ZnSe_001} the conductance for different
pinning positions in the gap is shown for Zn-terminated Fe/ZnSe(001)
with a 80 ML thick Schottky barrier. The minority-conductance peak at
$-0.2$~eV reflects the influence of the interface state.


Furthermore one can estimate the interface resistance under the
assumption of diffusive transport in the bulk, using Schep's
formula.\cite{SH97,SP00} Although this has been derived for metallic
multilayers, we have applied it to this case. Our results are
summarized in Table~\ref{tab.tunneling_results}. More details can be
found in Ref.~\onlinecite{Wunnicke03}.
\begin{table} [tb]
 \begin{ruledtabular}
 \begin{tabular} {llllll}
    & & $P$ (Zn) & $P$ (Se) & $P$ (Ga)& $P$ (As) \\
  \hline
  $R_\square$ $[\Omega m^2]$ & (8 ML) & $1 \cdot 10^{-10}$ & $8 \cdot 10^{-11}$ &
         $ 2 \cdot 10^{-10}$  & $2 \cdot 10^{-10}$   \\
  $P$ & (8 ML) & 96\% & 96\% & 99.8\%  & 99.8\%  \\
  \hline
  $R_\square$ $[\Omega m^2]$ & (80 ML) & $5 \cdot 10^{-5}$  & $ 7 \cdot 10^{-5}$  & 
        $ 2\cdot 10^{-9}$ & $ 4 \cdot 10^{-9}$ \\
  $P$ & (80 ML) & 77\% & 97\% & 99.2\% & 98.0\% \\
 \end{tabular}
 \end{ruledtabular}
 \caption{Interface resistance $R_I$ and spin polarization $P$ for two Schottky
   barrier thicknesses and all four terminations in Fe/SC(001).
   Results are obtained by integration over the whole 2DBZ. A
   thickness of 80 ML corresponds to 115\AA.}
 \label{tab.tunneling_results}
\end{table}

\section{Higher bias voltage \label{Sec:Bias}}

Up to now it was always assumed that the applied bias voltage is so
small that only states at $E_F$ carry current. But experimentally
especially for thicker barriers, a bias voltage in the order of 1V or
less is used. \cite{HJ02} This relatively high bias in experiments
with an optical detection of the spin polarization is needed to gain a
reasonable signal-to-noise ratio of the emitted light. It is
believed that the bias can be reduced in experiments with thinner
tunneling barriers or with an electrical detection of the
polarization, since then a much smaller current is needed. If a
non-zero bias is applied, one must integrate the transmission over the
appropriate energy range, and the Landauer formula reads\cite{CV97}
\begin{eqnarray}
\lefteqn{G = \frac{1}{eU} \int \limits_0^{-e U} dE} \nonumber  \\
&& {}\frac{e^2}{h} \frac{1}{A_\mathrm{2DBZ}} 
\sum \limits_{\sigma,\nu,\nu'} 
\int \limits_{\mathrm{2DBZ}} d^2k_\parallel \ 
T^\sigma_{\nu,\nu'} (\kp,E_F + E) 
\label{eq.landauer_with_bias}
\end{eqnarray}
with $T^\sigma_{\nu,\nu'} (\kp,E_F+E)$ being the transmission
probability at the energy $E_F + E$. This equation gives the standard
Landauer formula in the case of small bias voltages $U$ in a first
approximation, under the assumption that the electronic structure is
unaffected by the applied voltage.  

Here we show that also for a non-zero bias and a tunneling barrier of
moderate thickness the $\Gbar$ point is of main importance and a high
spin polarization is obtained. It has been shown \cite{MP00} that in
tunneling the smallest damping factor $\kappa$ is located at the
$\Gbar$ point in the energy gap for ZnSe and GaAs (and for other
direct-gap semiconductors), so that this point plays the major role
for thick tunneling barriers. Also at this point the symmetry enforced
spin polarization is the highest, as shown in the $\kp$-resolved hot
injection process. In the calculations the Fermi energy in the bulk SC
is assumed to be 10~meV above the conduction band minimum. For the Zn
termination an abrupt interface is taken for simplicity. Due to the
high bias voltage a $\kp$ integration over the whole 2DBZ is
performed, because also states away from the $\Gbar$ point in 2DBZ are
populated and carry the current.

\begin{figure} [tb]
  \begin{center}
    \includegraphics[width=3.0in,angle=0]{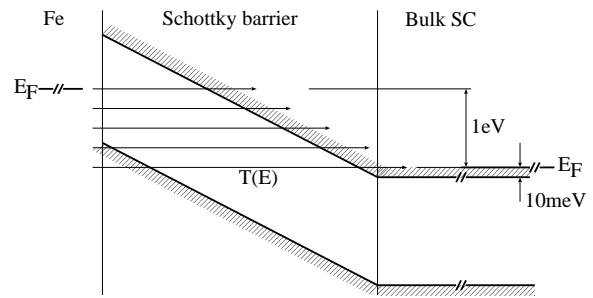}
    \caption{Sketch of the position of the energy bands and the
      energies in the injection process with an applied bias voltage
      and a tunneling barrier. The bias voltage is 1V.}
    \label{bias_sketch}
  \end{center}
\end{figure}

The position of the energy bands and of the Fermi
energy in the Fe and SC half-spaces
 are sketched in  Fig.~\ref{bias_sketch}.  
Already from this simple sketch it is qualitatively clear that the
conductance at the upper energy range (at the Fermi energy of the Fe
half-space) is the most important one, because there the effective
tunneling barrier is the smallest. In Table~\ref{tab.bias_results} the
spin polarizations are listed for a Fe/ZnSe(001) junction with an
applied voltage of 1V and a tunneling barrier of 8 ML and 80 ML
thickness. 
\begin{table} [tb]
 \begin{ruledtabular}
 \begin{tabular} {llllll}
   Barrier thickness (ML)  & $P$ (Zn) & $P$ (Se) \\
  \hline
  8  & 63.6\%  & 61.4\% \\
  \hline
  80  & 43.8\%  & 87.5\% \\
 \end{tabular}
 \end{ruledtabular}
 \caption{Current spin polarization in Fe/ZnSe(001) junctions with an
   applied bias voltage of 1V. The results are obtained by an
   integration over the whole 2DBZ. In the Bulk ZnSe(001) the Fermi
   energy lies 10meV in the conduction band.}
 \label{tab.bias_results}
\end{table}
In case of a Zn termination it is important that slightly below
the Fermi energy of the Fe half-space the resonant interface state
contributes largely to the minority conductance. This gives a
relatively low and a more or less constant spin
polarization with thicker tunneling barriers. In the Se
terminated interface the resonant interface state lies lower in
energy, where the effective tunneling barrier is thicker than for the
Zn terminated interface. Thus the resonant interface state has a much
smaller influence on the minority conductance. With larger barrier
thickness the high spin polarization is more or less restored because
only states at $\Gbar$, having the smallest decay parameter are
important and the symmetry arguments apply again.


\section{Summary and conclusions}\label{summary}

We have reported first-principles calculations on spin injection
from Fe into GaAs and ZnSe through the (001), (111), and (110)
interfaces. The electronic transport has been assumed to be in the
ballistic regime, and the interfaces have been assumed to have
two-dimensional periodicity, so that $\kp$ is conserved during
scattering at the interface. Then, in a Landauer-B\"uttiker approach,
the conductance is determined by the transmission probability summed
for all $\kp$. Under these assumptions, we have considered hot and
thermal injection --- the latter also in the presence of a Schottky
barrier --- and approached the higher bias regime. We have reached the
following conclusions.

(i) The spin polarization of the current is highest when most incoming
Fe Bloch states of one spin direction are totally reflected due to
symmetry mismatch with the SC conduction band states --- we call this
effect ``symmetry-enforced spin polarization''. This is the case for
the (001) interface, where minority-spin electrons for $\kp =0$
practically cannot be transmitted. The lower symmetry of the (111) and
(110) interface does not lead to such a selection rule.  Even for the
(001) interface, in the case of hot injection, above an energy
threshold where the Fe minority $\Delta_1$ band starts, the
minority-spin current increases rapidly even at $\kp=0$, being then
symmetry-allowed.

(ii) For the symmetry-enforced spin polarization to be realised, the
injection must take place close to the conduction band edge, so that
only states near $\kp=0$ are relevant; these have a well-defined
and suitable symmetry properties. Also, the interface should be as
ordered as possible, otherwise $\kp$ is not conserved and incoming
minority-spin Fe states from all $\kp$ can scatter into the SC
conduction band. Then the spin polarization will decrease.

(iii) The case of thermal injection ({\it i.e.} exactly at $E_F$), has
been considered for the (001) interface with and without a Schottky
barrier. The symmetry concepts still hold, since the least-decaying
complex band of the tunneling barrier has the same symmetry properties
as the conduction band (both are of $\Delta_1$ character). We have
seen that resonant interface states existing in the vicinity $E_F$ for
minority spin can be important for the current spin polarization. Even
in the case of a Schottky barrier they can provide a resonant
tunneling channel for the minority-spin electrons decreasing the spin
injection effect.

(iv) In the case of a higher bias, a wider energy range must be
considered, so that states with $\kp\neq 0$ as well as resonant states
close enough to $E_F$ become relevant. Then the spin polarization
decreases, but spin injection is still achieved.

It can be argued that a well ordered interface, necessary for our
symmetry selection rule, is difficult to realize experimentally.
However, recent successful attempts in increasing the spin injection
efficiency have been reported, accompanied by an improvement of the
quality of the interface.\cite{HE03} Therefore, we are optimistic that
our work will motivate further research in this direction.

\begin{acknowledgments}

The authors thank Arne Brataas for helpful comments. This work was
supported by the RT Network \textit{Computational Magnetoelectronics}
(Contract RTN1-1999-00145) of the European Commission.

\end{acknowledgments}

\appendix


\section{Geometry of the reduced 2DBZ in (001), (111), and (110)}

\begin{figure} [tb]
  \begin{center}
    \includegraphics[width=2.4in,angle=0]{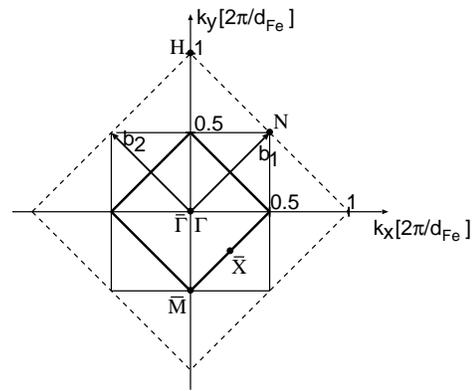}
    \caption{2DBZ for Fe(001) adapted
      to the Fe/SC(001) interface (thick line). For comparison the
      2DBZ for a free Fe(001) surface (thin solid
      line) and the (001) cut through the bulk Brillouin zone (thin
      dashed line) are also shown. The $k_z$ vector along the [001]
      direction varies between $\pm 2\pi/d_\mathrm{Fe}$.}
    \label{bz_Fe_001}
  \end{center}
\end{figure}

First we investigate the backfolded bands for the Fe(001)
half-space. In Fig.~\ref{bz_Fe_001} the 2DBZ and the cut perpendicular
to the [001] direction of the bulk Brillouin zone are shown.
At the $\Gbar$ point two bands are available: the not backfolded band
corresponding to the bulk band along $\Gamma$-H, {\it i.e.}, the $\Delta$
direction. This direction has a $C_{4v}$ symmetry. Also one band is
backfolded by applying once the reciprocal surface lattice vectors
$\mathbf{b_1}$ or $\mathbf{b_2}$, corresponding to the bulk band along
N-P-N, {\it i.e.}, the D direction.

\begin{figure} [tb]
  \begin{center}
    \includegraphics[width=2.3in,angle=0]{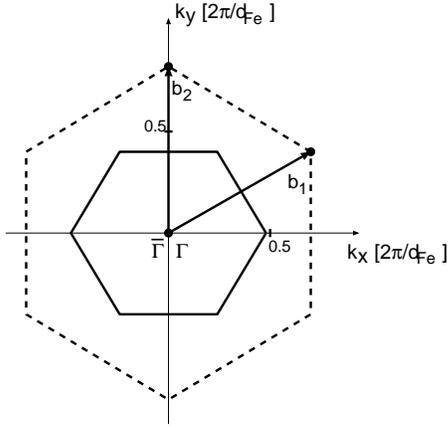}
    \caption{(111) cut through the bulk Brillouin zone (dashed line)
      and 2DBZ (solid line) of 
      Fe(111). The $k_z$ vector varies in the 2DBZ between $\pm \sqrt{3}
       \times 2\pi/d_\mathrm{Fe}$.}
    \label{bz_Fe_111}
  \end{center}
\end{figure}

Next we discuss the (111) case. In Fig.~\ref{bz_Fe_111} the (111) cut
through the bulk Brillouin zone and the 2DBZ are shown. As for the
(001) orientation additional bands are obtained by backfolding to the
$\Gbar$ point due to the large lattice constant in the Fe half-space.
The not backfolded band at the $\Gbar$ point corresponds to the bulk
band along the $\Gamma$-P-H high symmetry line, because in
the 2DBZ the $k_z$ vector extends from $-\sqrt{3}$ to $\sqrt{3}$ in
units of $2\pi/d_{Fe}$. By applying $\mathbf{b}_1$ 
or $\mathbf{b}_2$, together one additional band is backfolded to the
$\Gbar$ point. It corresponds to the bands of the same
$\Gamma$-P-H high symmetry line as the not backfolded band. 

\begin{figure} [tb]
  \begin{center}
    \includegraphics[width=1.8in,angle=270]{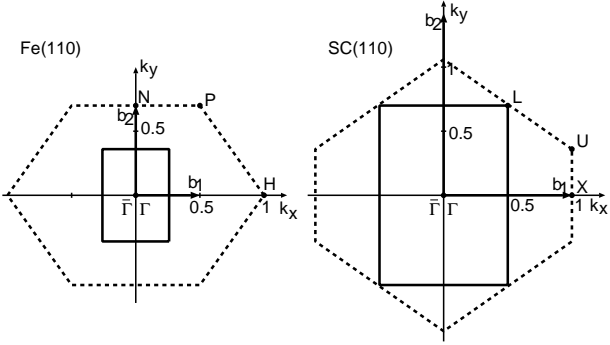}
    \caption{2DBZ (solid lines) and the (110) cut through the bulk
      Brillouin zone (dashed lines) for Fe(110) (left panel) and
      SC(110) (right panel). Values are given in $2\pi /d$,
      with $d$ being $d_\mathrm{Fe}$ and $d_\mathrm{SC}$ for the left
      and right panel, respectively.}
    \label{bz_110}
  \end{center}
\end{figure}

The 2DBZs of Fe and SC(110) and the corresponding (110) cut through
the bulk Brillouin zones are shown in Fig.~\ref{bz_110}.
In Fe(110) there are four backfolded bands by applying the reciprocal
surface lattice vectors $\mathbf{b}_1$ and 
$\mathbf{b}_2$.  The not backfolded band corresponds to the $\Gamma$-N
bulk band ($\Sigma$ direction). By applying the $\mathbf{b}_1$
reciprocal lattice vector once, 
one ends up with a backfolded band that is not along a high symmetry
line. More precisely, it is along the shortest line that connects two
P points in the bulk Brillouin zone. Written in components of the
reciprocal bulk lattice vectors, this backfolded band corresponds to
the bulk band along
\[  
\left(\begin{array}{c}-1/4\\-1/4\\ \phantom{-}1/2\end{array}\right)
  \frac{2\pi}{d_\mathrm{Fe}} \quad \rightarrow \quad
  \left(\begin{array}{c}1/4\\1/4\\1/2\end{array}\right) 
  \frac{2\pi}{d_\mathrm{Fe}} .
\]
If both reciprocal lattice vectors, $\mathbf{b}_1$ 
and $\mathbf{b}_2$, are applied, the backfolded band is parallel to
the one, if only $\mathbf{b}_1$ is applied, but shifted in the (110)
direction. Written in components of the reciprocal bulk lattice
vectors, it corresponds to
\[  
\left(\begin{array}{c}1/4\\1/4\\1/2\end{array}\right)
  \frac{2\pi}{d_\mathrm{Fe}} \quad \rightarrow \quad
  \left(\begin{array}{c}3/4\\3/4\\1/2\end{array}\right) 
  \frac{2\pi}{d_\mathrm{Fe}}.
\]
The next backfolded band is obtained by applying $\mathbf{b}_2$ once
that corresponds to the bulk band between the high symmetry points
N-H.  In summary due to the large two-dimensional surface unit cell
and small 2DBZ there are four bands available at the $\Gbar$ point for
the (110) orientation in the Fe half-space including two bands
corresponding \emph{not} to a bulk band along a high-symmetry line.
For these bands the only symmetry operation is the identity one, so
that they are allowed to couple to any SC bands due to symmetry.

In the SC (110) SBZ the not backfolded band corresponds to the bulk
high symmetry line $\Gamma$-$K$ ($\Sigma$-direction). By applying
$\mathbf{b}_1$ one additional band is backfolded to the $\Gbar$ point,
exactly on the not backfolded band. So the maximum conductance can be
2~$e^2/h$ per spin channel.


\end{document}